# *Modal expansions in dispersive material systems with application to quantum optics and topological photonics*

*Mário G. Silveirinha*[*]

*(1) University of Lisbon–Instituto Superior Técnico and Instituto de Telecomunicações, Avenida Rovisco Pais, 1, 1049-001 Lisboa, Portugal, mario.silveirinha@co.it.pt*

## I. Introduction

The response of many physical systems is often described by a Hermitian operator. For such systems the spectral theorem guarantees that an arbitrary state vector can be written in terms of the normal modes of oscillation. Thus, modal expansions are ubiquitous in many branches of physics and engineering. In particular, they enable one to construct formal solutions of many linear partial-differential equations of mathematical physics and to interpret and comprehend more easily the associated physical effects. Here, we are particularly interested in light phenomena. Modal expansions are widely used in such a context, for example, to obtain the solutions of radiation problems in waveguides, cavities or photonic crystals [1, 2, 3], to derive analytical expressions for Green's functions [1, 2, 3], in quantum optics [4], and in many analytical developments, just to name some of the most common applications.

In the literature the use of modal expansions is typically restricted to dispersionless structures, i.e., to the case wherein the relevant materials respond instantaneously (without any time delay) to an external excitation. One of the reasons is that from the Kramers-Kronig relations the material dispersion is typically accompanied by absorption and for lossy structures the Maxwell's equations cannot be reduced to a Hermitian operator problem. Crucially, the difficulties remain even in the absence (or for vanishingly small) material loss. Indeed, for dispersive systems the material response is written in terms of a convolution in time, and in general it is not obvious how to reformulate the system dynamics in terms of a standard (Schrödinger-type) Hermitian operator problem.

---

[*] To whom correspondence should be addressed: E-mail: *mario.silveirinha@co.it.pt*

The objective of this Chapter is to highlight that for lossless material platforms formed by arbitrary inhomogeneous bianisotropic and possibly nonreciprocal materials the natural modes of oscillation form, indeed, a complete set of expansion functions. Based on our recent work [5-6], it is proven that the Maxwell's equations in dispersive systems can always be reduced to a generalized dynamical problem whose time evolution is described by a Hermitian operator. The effects of material dispersion are taken into account by introducing additional variables that may model the internal degrees of freedom of the material. With such a result, we construct formal expansions of the electromagnetic field in terms of the normal modes, and in particular it is highlighted that the modal expansion coefficients are not unique. The developed theory is used to obtain a modal expansion of the system Green's function.

Furthermore, based on the equivalence between the dispersive Maxwell's equations and the generalized dynamical problem we discuss the definitions of energy and momentum in a dispersive system. In particular, it is shown that for time-harmonic solutions the energy and the total momentum can be written exclusively in terms of the electromagnetic degrees of freedom.

In addition, we illustrate the application of the proposed formalism in the context of *i)* topological photonics and *ii)* quantum optics. It is shown that the notions of Berry potential and Berry curvature can be generalized in a straightforward manner to a generic dispersive system, and in particular our theory enables the topological classification of a generic bianisotropic and nonreciprocal electromagnetic continuum. Moreover, we develop a simple and intuitive theory for the quantization of the electromagnetic fields in dispersive material systems and for the calculation of the quantum field correlations. Our approach yields results that agree with those obtained with other more sophisticated approaches and with the fluctuation-dissipation theorem.

## II. Electrodynamics of dispersive media

In this chapter, we consider frequency dispersive media with vanishing (or weak) material loss. The Maxwell's equations in the time domain can be written in a compact manner as

$$\hat{N} \cdot \mathbf{f} = i \left[ \frac{\partial \mathbf{g}}{\partial t} + \mathbf{j} \right]. \tag{1}$$

We adopt six-vector notations with $\mathbf{f} = \begin{pmatrix} \mathbf{E} & \mathbf{H} \end{pmatrix}^T$, $\mathbf{g} = \begin{pmatrix} \mathbf{D} & \mathbf{B} \end{pmatrix}^T$, $\mathbf{j} = \begin{pmatrix} \mathbf{j}_e & \mathbf{j}_m \end{pmatrix}^T$, where $\mathbf{E}, \mathbf{H}$ are the electric and magnetic fields, $\mathbf{D}, \mathbf{B}$ are the electric displacement and the induction fields, $\mathbf{j}_e, \mathbf{j}_m$ are the electric and magnetic current densities. The superscript $T$ denotes the transposition operation. The differential operator $\hat{N}$ is defined as:

$$\hat{N} = \begin{pmatrix} \mathbf{0} & i\nabla \times \mathbf{1}_{3\times 3} \\ -i\nabla \times \mathbf{1}_{3\times 3} & \mathbf{0} \end{pmatrix}, \tag{2}$$

where $\mathbf{1}_{3\times3}$ is the identity matrix of dimension three. The relation between the $\mathbf{f}$ and $\mathbf{g}$ vector fields is determined by the material constitutive relations. For dispersive linear media, the electromagnetic response is not instantaneous, and thereby generally $\mathbf{f}$ and $\mathbf{g}$ are linked by a convolution in time. In the spectral (frequency) domain, $\mathbf{f}$ and $\mathbf{g}$ are simply related by a multiplication operator $\mathbf{M}$:

$$\mathbf{g}(\mathbf{r},\omega) = \mathbf{M}(\mathbf{r},\omega) \cdot \mathbf{f}(\mathbf{r},\omega), \quad \text{with} \quad \mathbf{M} = \begin{pmatrix} \varepsilon_0 \overline{\varepsilon} & \frac{1}{c}\overline{\xi} \\ \frac{1}{c}\overline{\zeta} & \mu_0 \overline{\mu} \end{pmatrix}. \tag{3}$$

The components of the 6×6 material matrix $\mathbf{M}$ are the relative permittivity $\overline{\varepsilon}$, the relative permeability $\overline{\mu}$, and the tensors $\overline{\xi}, \overline{\zeta}$ associated with magneto-electric effects (e.g., chirality or more generally bianisotropy) [7, 8]. Since the electromagnetic fields are real-valued in the time-domain the material response must satisfy [9, 10]:

$$\mathbf{M}(\omega) = \mathbf{M}^*(-\omega^*), \tag{4}$$

where the superscript "*" denotes the complex conjugation operation. For a lossless medium the material matrix must be Hermitian for $\omega$ real-valued [8, 11]:

$$\mathbf{M}(\omega) = \mathbf{M}^\dagger(\omega). \tag{5}$$

The superscript "†" denotes the conjugate transpose matrix. Moreover, the medium response must also guarantee that $\frac{\partial}{\partial \omega}[\omega \mathbf{M}(\omega)]$ is a positive definite matrix,

$$\frac{\partial}{\partial \omega}[\omega \mathbf{M}(\omega)] > 0, \tag{6}$$

so that the time-averaged stored energy density is nonnegative [8].

Due to the causality of the material response, $\mathbf{M}$ must be an analytic function of $\omega$ in the upper-half frequency plane ($\mathrm{Im}\{\omega\} > 0$) [9, 10]. It is further assumed that $\mathbf{M}$ is meromorphic in the complex plane so that all the singular points are poles. For a lossless system all the poles are in the real-frequency axis. Let $\mathbf{M}_\infty = \lim_{\omega \to \infty} \mathbf{M}(\omega)$ represent the asymptotic high-frequency response of the material. Usually, $\mathbf{M}_\infty$ reduces to the material matrix of the vacuum with diagonal elements $\varepsilon_0, \mu_0$. Using the Cauchy theorem and Eqs. (4)-(6) it is straightforward to prove that for $\omega$ real-valued $\mathbf{M}$ has the following partial-fraction expansion [5-6]:

$$\mathbf{M}(\omega) = \mathbf{M}_\infty - \sum_\alpha \frac{\mathrm{sgn}(\omega_{p,\alpha})}{\omega - \omega_{p,\alpha}} \mathbf{A}_\alpha^2 . \tag{7}$$

Here, $\mathrm{sgn} = \pm$ is the sign function, $\omega_{p,\alpha}$ are the (real-valued) poles of $\mathbf{M}$, $(\mathrm{Res}\mathbf{M})_\alpha$ are the corresponding residues, and $\mathbf{A}_\alpha = \left[-\mathrm{sgn}(\omega_{p,\alpha})(\mathrm{Res}\mathbf{M})_\alpha\right]^{1/2}$. It can be shown that $\mathbf{A}_\alpha$ is a positive (semi-)definite matrix ($\mathbf{A}_\alpha \geq 0$).

### III. Hermitian formulation in the time domain

In frequency dispersive structures it is not obvious how to obtain modal expansions of the electromagnetic fields. In the following, it is proven that in a rather general context it is possible to introduce additional variables that describe the dynamics of the medium internal degrees of freedom responsible for the non-instantaneous response. Furthermore, we show that for lossless systems the electromagnetic problem can be formulated as Schrödinger-type (eventually inhomogeneous) equation with a Hermitian time evolution operator.

As a starting point, we rewrite the Maxwell's equations (1) in the spectral (frequency) domain:

$$\hat{N} \cdot \mathbf{f} = \omega \mathbf{M}(\mathbf{r}, \omega) \cdot \mathbf{f} + i\mathbf{j}. \tag{8}$$

Next, we introduce the auxiliary fields:

$$\mathbf{Q}^{(\alpha)} = \frac{|\omega_{p,\alpha}|^{1/2}}{(\omega - \omega_{p,\alpha})} \mathbf{A}_\alpha \cdot \mathbf{f}, \qquad \alpha = 1, 2, ...., N_p \tag{9}$$

where $\omega_{p,\alpha}$ and $\mathbf{A}_\alpha$ are defined as in Sect. II, and in general depend on the spatial coordinates $\mathbf{r}$. Note that $\omega_{p,\alpha}$ and $\mathbf{A}_\alpha$ are typically discontinuous functions of $\mathbf{r}$, i.e. depend on the considered material. Furthermore, the number of poles typically depends on the material region. The number of auxiliary fields ($N_p$) is determined by the material with the larger number of poles. To minimize the bookkeeping, it is convenient to define $\mathbf{Q}^{(\alpha)}$ over all space. Thus, we take $\mathbf{A}_\alpha(\mathbf{r}) = 0$ for $\alpha > N_{p,i}$ and $\mathbf{r}$ in the $i$-th material region. Here, $N_{p,i}$ is the number of poles of the $i$-th material. From the definition of $\mathbf{Q}^{(\alpha)}$, it is clear that:

$$\omega \mathbf{Q}^{(\alpha)} = \omega_{p,\alpha} \mathbf{Q}^{(\alpha)} + |\omega_{p,\alpha}|^{1/2} \mathbf{A}_\alpha \cdot \mathbf{f}. \tag{10}$$

On the other hand, from the partial-fraction decomposition of the material matrix (7) we get:

$$\mathbf{g} = \mathbf{M} \cdot \mathbf{f} = \mathbf{M}_\infty \cdot \mathbf{f} - \sum_\alpha \operatorname{sgn}(\omega_{p,\alpha}) \frac{1}{|\omega_{p,\alpha}|^{1/2}} \mathbf{A}_\alpha \cdot \mathbf{Q}^{(\alpha)}. \tag{11}$$

Then, with the help of Eq. (10) it is possible to write $\omega \mathbf{g} = \omega \mathbf{M}_\infty \cdot \mathbf{f} - \sum_\alpha |\omega_{p,\alpha}|^{1/2} \mathbf{A}_\alpha \cdot \mathbf{Q}^{(\alpha)} - \sum_\alpha \operatorname{sgn}(\omega_{p,\alpha}) \mathbf{A}_\alpha^2 \cdot \mathbf{f}$. Substituting this formula into the Maxwell's equations (8), and calculating the inverse Fourier transform in time of the resulting equation and of Eq. (10), one finds that the Maxwell's equations (1) are equivalent to a first-order in time partial differential system of the form $\hat{L} \cdot \mathbf{Q} = i \frac{\partial}{\partial t} \mathbf{M}_g \cdot \mathbf{Q} + i \mathbf{j}_g$ with:

$$\underbrace{\begin{pmatrix} \hat{N} + \sum_\alpha \text{sgn}(\omega_{p,\alpha}) \mathbf{A}_\alpha^2 & |\omega_{p,1}|^{1/2} \mathbf{A}_1 & |\omega_{p,2}|^{1/2} \mathbf{A}_2 & \cdots \\ |\omega_{p,1}|^{1/2} \mathbf{A}_1 & \omega_{p,1} \mathbf{1} & 0 & \cdots \\ |\omega_{p,2}|^{1/2} \mathbf{A}_2 & 0 & \omega_{p,2} \mathbf{1} & \cdots \\ \cdots & \cdots & \cdots & \cdots \end{pmatrix}}_{\hat{L}} \cdot \underbrace{\begin{pmatrix} \mathbf{f} \\ \mathbf{Q}^{(1)} \\ \mathbf{Q}^{(2)} \\ \cdots \end{pmatrix}}_{\mathbf{Q}} = i \frac{\partial}{\partial t} \underbrace{\begin{pmatrix} \mathbf{M}_\infty & 0 & 0 & \cdots \\ 0 & 1 & 0 & \cdots \\ 0 & 0 & 1 & \cdots \\ \cdots & \cdots & \cdots & \cdots \end{pmatrix}}_{\mathbf{M}_g} \cdot \begin{pmatrix} \mathbf{f} \\ \mathbf{Q}^{(1)} \\ \mathbf{Q}^{(2)} \\ \cdots \end{pmatrix} + i \underbrace{\begin{pmatrix} \mathbf{j} \\ 0 \\ 0 \\ \cdots \end{pmatrix}}_{\mathbf{j}_g}$$

(12)

where $\mathbf{1} \equiv \mathbf{1}_{6 \times 6}$. We introduced the generalized state vector $\mathbf{Q} = \begin{pmatrix} \mathbf{f} & \mathbf{Q}^{(1)} & \cdots & \mathbf{Q}^{(\alpha)} & \cdots \end{pmatrix}^T$, which describes the dynamics of the electromagnetic field and of the internal degrees of freedom of the material response (see also [12-16]). Note that $\hat{L}$ is a differential operator and that in an inhomogeneous material platform both $\hat{L}$ and $\mathbf{M}_g$ depend on the spatial coordinates. The operators $\hat{L}$ and $\mathbf{M}_g$ are Hermitian with respect to the canonical inner product because $\mathbf{A}_\alpha$ and $\mathbf{M}_\infty$ are Hermitian matrices. This property implies that the operator $\hat{H}_{cl}(\mathbf{r}, -i\nabla) = \mathbf{M}_g^{-1} \cdot \hat{L}$ is Hermitian with respect to the weighted inner product:

$$\langle \mathbf{Q}_B | \mathbf{Q}_A \rangle \equiv \int_V \frac{1}{2} \mathbf{Q}_B^* \cdot \mathbf{M}_g(\mathbf{r}) \cdot \mathbf{Q}_A d^3\mathbf{r}. \tag{13}$$

Here, $V$ represents the domain of interest and it is assumed that the boundary conditions (e.g., periodic boundary conditions) ensure that $\hat{H}_{cl}$ is indeed Hermitian. Hence, in the absence of an external excitation ($\mathbf{j}_g = 0$), the generalized system (12) is equivalent to $\hat{H}_{cl} \cdot \mathbf{Q} = i \frac{\partial}{\partial t} \mathbf{Q}$, which is analogous to the Schrödinger equation with $\hbar = 1$.

In a conventional time-evolution problem one needs to specify the value of the fields at the initial time instant, let us say $t=0$. Clearly, in a dispersive material system the solution of the time-evolution problem requires the knowledge of the initial time electromagnetic fields $\mathbf{f}$ as well as the initial time additional variables $\mathbf{Q}^{(\alpha)}$.

## IV. Poynting theorem and stored energy

Straightforward manipulations of the dynamical equation $\hat{L} \cdot \mathbf{Q} = i \frac{\partial}{\partial t} \mathbf{M}_g \cdot \mathbf{Q} + i \mathbf{j}_g$ [Eq. (12)] shows that the state vector $\mathbf{Q}$ satisfies the following generalized Poynting theorem:

$$\nabla \cdot \mathbf{S} + \partial_t W = -\frac{1}{2}\left(\mathbf{Q}^* \cdot \mathbf{j}_g + \mathbf{Q} \cdot \mathbf{j}_g^*\right). \tag{14}$$

Here, $\mathbf{S}(\mathbf{r},t) = \frac{1}{2}\left(\mathbf{E} \times \mathbf{H}^* + \mathbf{E}^* \times \mathbf{H}\right)$ is the instantaneous Poynting vector and $W$ is the instantaneous stored energy density:

$$W(\mathbf{r},t) = \frac{1}{2}\mathbf{Q}^* \cdot \mathbf{M}_g \cdot \mathbf{Q}. \tag{15}$$

Thus, the stored energy density is determined by a weighted scalar product. Note that for completeness we consider complex-valued solutions of the dynamical equation (12) in the time-domain.

Importantly, for time-harmonic solutions of (12) the inner-product can be written simply in terms of the electromagnetic fields. To demonstrate this result, we consider two generic solutions, $\mathbf{Q}_A(\mathbf{r},t) = \tilde{\mathbf{Q}}_A(\mathbf{r})e^{-i\omega_A t}$ and $\mathbf{Q}_B(\mathbf{r},t) = \tilde{\mathbf{Q}}_B(\mathbf{r})e^{-i\omega_B t}$ of the generalized problem (12), with a time dependence of the form $e^{-i\omega t}$ and envelopes $\tilde{\mathbf{Q}}_A$ and $\tilde{\mathbf{Q}}_B$. The two oscillation frequencies $\omega_A$ and $\omega_B$ may be different. The state vectors $\mathbf{Q}_A$ and $\mathbf{Q}_B$ are not necessarily natural modes of the system, and may be associated with some external current excitations, $\mathbf{j}_A$ and $\mathbf{j}_B$, respectively. However, it is required that the generalized excitation vector $\mathbf{j}_g$ in Eq. (12) is of the form $\mathbf{j}_g = \begin{pmatrix} \mathbf{j} & 0 & \ldots & 0 & \ldots \end{pmatrix}^T$, i.e., the excitation can only act on the electromagnetic degrees of freedom so that Eq. (9) holds.

Let $\mathbf{f}_A(\mathbf{r},t) = \mathbf{F}_A(\mathbf{r})e^{-i\omega_A t}$ and $\mathbf{f}_B(\mathbf{r},t) = \mathbf{F}_B(\mathbf{r})e^{-i\omega_B t}$ be the corresponding solutions of the time-harmonic Maxwell's equations (8). Using Eq. (9) one finds that the scalar product of $\tilde{\mathbf{Q}}_A$ and $\tilde{\mathbf{Q}}_B$ is

$$\frac{1}{2}\tilde{\mathbf{Q}}_B^* \cdot \mathbf{M}_g(\mathbf{r}) \cdot \tilde{\mathbf{Q}}_A = \frac{1}{2}\left[\mathbf{F}_B^* \cdot \mathbf{M}_\infty \cdot \mathbf{F}_A + \mathbf{F}_B^* \cdot \sum_\alpha \mathbf{A}_\alpha \frac{|\omega_{p,\alpha}|^{1/2}}{(\omega_B - \omega_{p,\alpha})} \cdot \frac{|\omega_{p,\alpha}|^{1/2}}{(\omega_A - \omega_{p,\alpha})} \mathbf{A}_\alpha \cdot \mathbf{F}_A\right]. \quad (16)$$

With the help of Eq. (7) the above result can be written in a compact manner as:

$$\frac{1}{2}\tilde{\mathbf{Q}}_B^* \cdot \mathbf{M}_g(\mathbf{r}) \cdot \tilde{\mathbf{Q}}_A = \frac{1}{2}\mathbf{F}_B^* \cdot \left[\frac{\omega_B \mathbf{M}(\mathbf{r},\omega_B) - \omega_A \mathbf{M}(\mathbf{r},\omega_A)}{\omega_B - \omega_A}\right] \cdot \mathbf{F}_A. \quad (17)$$

Thus, the inner product of $\tilde{\mathbf{Q}}_A$ and $\tilde{\mathbf{Q}}_B$ can be expressed simply in terms of the "electromagnetic envelopes", $\mathbf{F}_A$ and $\mathbf{F}_B$ [5-6]:

$$\langle \tilde{\mathbf{Q}}_B | \tilde{\mathbf{Q}}_A \rangle = \frac{1}{2}\int_V d^3\mathbf{r}\, \mathbf{F}_B^*(\mathbf{r}) \cdot \left[\frac{\omega_B \mathbf{M}(\mathbf{r},\omega_B) - \omega_A \mathbf{M}(\mathbf{r},\omega_A)}{\omega_B - \omega_A}\right] \cdot \mathbf{F}_A(\mathbf{r}). \quad (18)$$

Clearly, when $\omega_B = \omega_A \equiv \omega$ the right-hand side of Eq. (17) has a singular behavior. The singularity can be formally removed by taking the limit $\omega_B \to \omega_A$, which is well defined because $\mathbf{Q}_A$ and $\mathbf{Q}_B$ are generally driven by external excitations $\mathbf{j}_A$ and $\mathbf{j}_B$ (alternatively, one may directly use Eq. (16) which has no singularities when $\omega_B = \omega_A \equiv \omega$). The outlined limit process yields

$$\frac{1}{2}\tilde{\mathbf{Q}}_B^* \cdot \mathbf{M}_g(\mathbf{r}) \cdot \tilde{\mathbf{Q}}_A = \frac{1}{2}\mathbf{f}_B^* \cdot \frac{\partial}{\partial \omega}\left[\omega \mathbf{M}(\mathbf{r},\omega)\right] \cdot \mathbf{f}_A. \tag{19}$$

In particular, it follows that the inner product of a time-harmonic solution of (12) with itself satisfies [5-6]:

$$\langle \mathbf{Q} | \mathbf{Q} \rangle = \frac{1}{2}\int_V d^3\mathbf{r}\, \mathbf{f}^*(\mathbf{r}) \cdot \frac{\partial}{\partial \omega}\left[\omega \mathbf{M}(\mathbf{r},\omega)\right] \cdot \mathbf{f}(\mathbf{r}). \tag{20}$$

Note that the right-hand side of the above expression represents the total energy in volume $V$, which due to Eq. (6) is always a positive quantity [2, 9, 10, 17, 18].

## V. Canonical momentum

Let us consider again a time-harmonic solution of the generalized system (12), $\mathbf{Q}(\mathbf{r},t) = \tilde{\mathbf{Q}}(\mathbf{r})e^{-i\omega t}$. For material structures invariant to translations along a given space direction, let us say the $x$-direction, we introduce a canonical (wave) momentum defined by:

$$\mathcal{P}_x = \frac{1}{\omega}\langle \mathbf{Q} | -i\partial_x | \mathbf{Q} \rangle. \tag{21}$$

The canonical momentum of dispersive isotropic media was first discussed in Refs. [19-20]. The definition (21) generalizes the theory of Refs. [19-20] to arbitrary bianisotropic and nonreciprocal layered structures. Note that $-i\partial_x$ is an Hermitian operator with respect to the weighted inner product (13) when $\partial_x \mathbf{M}_g = 0$, i.e., when the material parameters are independent of $x$. Thus, $\mathcal{P}_x$ is indeed a real-valued number. As shown next, $\mathcal{P}_x$ can be written as a function of the electromagnetic component of the state vector $(\mathbf{f}(\mathbf{r},t) = \mathbf{F}(\mathbf{r})e^{-i\omega t})$.

To prove this, we use Eq. (9) and take into account that the material parameters are independent of $x$ to obtain (compare with Eq. (16)):

$$\frac{1}{2}\mathbf{Q}^* \cdot \mathbf{M}_g(\mathbf{r}) \cdot (-i\partial_x)\mathbf{Q} = \frac{1}{2}\left[\mathbf{f}^* \cdot \mathbf{M}_\infty \cdot (-i\partial_x)\mathbf{f} + \mathbf{f}^* \cdot \sum_\alpha \frac{|\omega_{p,\alpha}|}{(\omega - \omega_{p,\alpha})^2} \mathbf{A}_\alpha^2 \cdot (-i\partial_x)\mathbf{f}\right]. \quad (22)$$

But from the partial fraction expansion (7) it follows that

$$\frac{1}{2}\mathbf{Q}^* \cdot \mathbf{M}_g(\mathbf{r}) \cdot (-i\partial_x)\mathbf{Q} = \frac{1}{2}\mathbf{f}^* \cdot \partial_\omega(\omega\mathbf{M}) \cdot (-i\partial_x)\mathbf{f}, \quad (23)$$

which implies that the canonical momentum is given by

$$\mathcal{P}_x = \frac{1}{2\omega}\int_V d^3\mathbf{r}\, \mathbf{f}^* \cdot \partial_\omega(\omega\mathbf{M}) \cdot (-i\partial_x)\mathbf{f}. \quad (24)$$

In particular, for a wave with a time-variation along $x$ of the form $e^{ik_x x}$ (the variation along $y$ and $z$ may be completely arbitrary; note that the material structure can be non-uniform along $y$ and $z$) it follows from (20) that

$$\mathcal{P}_x = \frac{k_x}{\omega}\mathcal{E}, \quad (25)$$

where $\mathcal{E}$ is the stored energy. This formula is consistent with the understanding that the field is formed by quanta with momenta $\hbar k_x$ because $\mathcal{E}/(\hbar\omega)$ may be understood as the number of quanta in the field [21, 22, 23].

It is interesting to connect the canonical momentum to the Minkowski momentum [22, 24]. To this end, we rewrite (24) as:

$$\mathcal{P}_x = \frac{1}{2\omega}\int_V d^3\mathbf{r}\left[\mathbf{g}^* \cdot (-i\partial_x)\mathbf{f} + \omega\mathbf{f}^* \cdot \partial_\omega\mathbf{M} \cdot (-i\partial_x)\mathbf{f}\right]. \quad (26)$$

where we introduced $\mathbf{g} = \mathbf{M}(\mathbf{r},\omega) \cdot \mathbf{f}$, so that $\mathbf{g} = \begin{pmatrix} \mathbf{D} & \mathbf{B} \end{pmatrix}^T$. Furthermore, we suppose in the rest of this section that $\mathbf{f}$ is a natural mode ($\mathbf{j} = 0$). In such a case, it is possible to show by direct manipulation of the frequency domain Maxwell's equations that (see Ref. [25, Ap. C]):

$$\mathbf{g}^* \cdot \partial_x \mathbf{f} = \nabla \cdot \left( \mathbf{D}^* \mathbf{E} \cdot \hat{\mathbf{x}} + \mathbf{B}^* \mathbf{H} \cdot \hat{\mathbf{x}} \right) + i\omega \hat{\mathbf{x}} \cdot \left( \mathbf{D}^* \times \mathbf{B} + \mathbf{D} \times \mathbf{B}^* \right). \tag{27}$$

Integrating both sides of the previous equation over the volume of interest and supposing that the boundary conditions (e.g., Bloch boundary conditions) ensure that the integral of the divergence term vanishes, it is found that:

$$\mathcal{P}_x = \int_V d^3\mathbf{r} \left[ \text{Re}\{\mathbf{D} \times \mathbf{B}^*\} \cdot \hat{\mathbf{x}} + \frac{1}{2} \mathbf{f}^* \cdot \partial_\omega \mathbf{M} \cdot (-i\partial_x) \mathbf{f} \right]. \tag{28}$$

Thus, for dispersionless media the canonical momentum is determined by the Minkowski momentum with momentum density $\mathbf{D} \times \mathbf{B}$ [22, 25]. In dispersive media the canonical momentum gains an extra term, as it was first shown in [19, 20, 26] for the case of isotropic dispersive dielectrics. As discussed in detail in Ref. [24], the canonical momentum must be understood as the *total* momentum of the system, and has contributions from the electromagnetic field as well as from the *kinetic* degrees of freedom of matter. Indeed, the light-matter interactions typically result in optical forces that set the medium into motion. For very massive bodies the center of mass velocity is insignificant, but the transferred kinetic momentum (i.e., the product of the total mass and of the center of mass velocity) cannot be neglected [24]. The wave momentum $\mathcal{P}_x$ is the sum of the light momentum and of the kinetic momentum of the medium [24]. It is underlined that this result applies *exclusively* to lossless media invariant to translations along *x* and to natural modes of the Maxwell's equations. Indeed, when the medium is inhomogeneous along *x*, typically the kinetic momentum transferred to the medium grows linearly with time in a steady state regime (with time variation $e^{-i\omega t}$) because of the stress created at material interfaces. Finally, we point out that for real-valued fields, $\mathbf{f}(\mathbf{r},t) = \text{Re}\{\mathbf{f}_\omega(\mathbf{r})e^{-i\omega t}\}$, the time-averaged canonical momentum is given by the right-hand side of Eq. (28) multiplied by an extra $1/2$ factor and with the replacement $\mathbf{f} \to \mathbf{f}_\omega$.

## VI. Modal expansions

Let us consider a generic state vector $\mathbf{Q}(\mathbf{r}) = \begin{pmatrix} \mathbf{f} & \mathbf{Q}^{(1)} & \ldots & \mathbf{Q}^{(\alpha)} & \ldots \end{pmatrix}^T$ defined over the volume of interest *V*. For example, *V* may correspond to a closed cavity. Since $\hat{H}_{cl}$ is Hermitian with respect to the weighted inner product (13), the spectral theorem guarantees that $\mathbf{Q}(\mathbf{r})$ can be expanded into a basis of eigenmodes of $\hat{H}_{cl}$, $\mathbf{Q}_1, \mathbf{Q}_2, \ldots$, (the natural modes of oscillation), which satisfy:

$$\hat{H}_{cl} \mathbf{Q}_n = \omega_n \mathbf{Q}_n, \tag{29}$$

where $\omega_n$ are the eigenfrequencies. Supposing that the eigenmodes are normalized such that $\langle \mathbf{Q}_n | \mathbf{Q}_m \rangle = \delta_{n,m}$ we have the modal expansion:

$$\mathbf{Q}(\mathbf{r}) = \sum_n \mathbf{Q}_n(\mathbf{r}) c_n \qquad \text{with} \qquad c_n = \langle \mathbf{Q}_n | \mathbf{Q} \rangle. \tag{30}$$

In particular, if we take $\mathbf{Q}(\mathbf{r}) \to \delta(\mathbf{r}-\mathbf{r}')\mathbf{1}_g$ with $\mathbf{1}_g$ the identity tensor (with the same dimension as $\mathbf{M}_g$) we obtain the completeness relation:

$$\delta(\mathbf{r}-\mathbf{r}')\mathbf{1}_g = \frac{1}{2}\sum_n \mathbf{Q}_n(\mathbf{r}) \otimes \mathbf{Q}_n^*(\mathbf{r}') \cdot \mathbf{M}_g(\mathbf{r}'), \tag{31}$$

where $\otimes$ represents the tensor product of two vectors.

-13-

Let $\mathbf{f}_n$ be the (envelope) electromagnetic field distribution associated with $\mathbf{Q}_n$, which thereby satisfies:

$$\hat{N} \cdot \mathbf{f}_n = \omega_n \mathbf{M}(\mathbf{r}, \omega_n) \cdot \mathbf{f}_n. \tag{32}$$

From Eqs. (18) and (20) the normalization condition $\langle \mathbf{Q}_n | \mathbf{Q}_m \rangle = \delta_{n,m}$ implies that:

$$\frac{1}{2}\int_V d^3\mathbf{r}\, \mathbf{f}_n^*(\mathbf{r}) \cdot \frac{\partial}{\partial \omega}\left[\omega \mathbf{M}(\mathbf{r},\omega)\right]_{\omega=\omega_n} \cdot \mathbf{f}_m(\mathbf{r}) = \delta_{n,m}, \quad \text{if } \omega_n = \omega_m. \tag{33a}$$

$$\frac{1}{2}\int_V d^3\mathbf{r}\, \mathbf{f}_n^*(\mathbf{r}) \cdot \left[\frac{\omega_n \mathbf{M}(\mathbf{r},\omega_n) - \omega_m \mathbf{M}(\mathbf{r},\omega_m)}{\omega_n - \omega_m}\right] \cdot \mathbf{f}_m(\mathbf{r}) = 0, \quad \text{if } \omega_n \neq \omega_m. \tag{33b}$$

Remarkably, the completeness relation (31) yields the following completeness relation for the electromagnetic modes [27]:

$$\delta(\mathbf{r}-\mathbf{r}')\mathbf{M}_\infty^{-1}(\mathbf{r}') = \frac{1}{2}\sum_n \mathbf{f}_n(\mathbf{r}) \otimes \mathbf{f}_n^*(\mathbf{r}'), \tag{34}$$

with the modes normalized as in (33). In particular, a generic electromagnetic field distribution $\mathbf{f}(\mathbf{r})$ has the modal expansion:

$$\mathbf{f}(\mathbf{r}) = \sum_n \mathbf{f}_n(\mathbf{r}) c_n, \quad \text{with} \quad c_n = \frac{1}{2}\int_V d^3\mathbf{r}'\, \mathbf{f}_n^*(\mathbf{r}') \cdot \mathbf{M}_\infty(\mathbf{r}') \cdot \mathbf{f}(\mathbf{r}'). \tag{35}$$

Importantly, in a dispersive medium the expansion coefficients $c_n$ are *not* unique. In other words, an identity of the type $\mathbf{f}(\mathbf{r}) = \sum_n \mathbf{f}_n(\mathbf{r}) c_n$ may be satisfied with different sets of coefficients $c_n$. For example, if we take $\mathbf{f}(\mathbf{r}) = \mathbf{f}_m(\mathbf{r})$ an obvious solution is $c_n = \delta_{n,m}$, which for a dispersive material clearly differs from the coefficients given by (35). The lack of uniqueness of the expansion coefficients follows from the fact that the electromagnetic component of $\mathbf{Q}$ describes only a subset of the degrees of freedom of the material. In contrast, the modal expansion (30) is unique.

Let us consider now that $\mathbf{f}(\mathbf{r})$ is some solution of the time-harmonic problem (8), and let $\mathbf{Q}$ be the corresponding solution of the generalized time-harmonic problem. Then, we may obtain a set of expansion coefficients for $\mathbf{f}(\mathbf{r})$ from Eqs. (18) and (30):

$$c_n = \frac{1}{2}\int_V d^3\mathbf{r}\, \mathbf{f}_n^*(\mathbf{r}) \cdot \left[\frac{\omega_n \mathbf{M}(\mathbf{r},\omega_n) - \omega \mathbf{M}(\mathbf{r},\omega)}{\omega_n - \omega}\right] \cdot \mathbf{f}(\mathbf{r})$$
$$= \frac{1}{2}\int_V d^3\mathbf{r}\, \frac{\left[\hat{N}\mathbf{f}_n\right]^* \cdot \mathbf{f} - \mathbf{f}_n^* \cdot \left(\hat{N}\mathbf{f} - i\mathbf{j}\right)}{\omega_n - \omega} \quad . \quad (36)$$

We used the fact that the material matrix is Hermitian and Eq. (32). But since the differential operator $\hat{N}$ is Hermitian with respect to the canonical inner product, we finally obtain:

$$\mathbf{f}(\mathbf{r}) = \sum_n \mathbf{f}_n(\mathbf{r}) c_n, \quad \text{with} \quad c_n = \frac{1}{2}\int_V d^3\mathbf{r}\, \frac{\mathbf{f}_n^* \cdot i\mathbf{j}}{\omega_n - \omega}. \quad (37)$$

We emphasize that the modal expansion (35) is universally valid, whereas the modal expansion (37) only holds when $\mathbf{f}$ is a time-harmonic solution of the Maxwell's equations with oscillation frequency $\omega$.

## VII. Green's function

We define the (causal) Green's function tensor in the time domain as the solution of [27]:

$$\left(\hat{N} - i\frac{\partial}{\partial t}\mathbf{M}\right) \cdot \overline{\mathbf{G}}(\mathbf{r},\mathbf{r}_0,t) = i\mathbf{1}\delta(\mathbf{r}-\mathbf{r}_0)\delta(t), \quad (38)$$

with $\mathbf{1} \equiv \mathbf{1}_{6\times 6}$. It is understood that the action of the material matrix on the time domain Green's function is given by a convolution. Evidently, in the spectral domain the Green's function satisfies (with a time-harmonic variation $e^{-i\omega t}$):

$$\hat{N} \cdot \overline{\mathbf{G}}(\mathbf{r},\mathbf{r}_0,\omega) = \omega \mathbf{M}(\mathbf{r},\omega) \cdot \overline{\mathbf{G}}(\mathbf{r},\mathbf{r}_0,\omega) + i\mathbf{1}\delta(\mathbf{r}-\mathbf{r}_0). \tag{39}$$

It is useful to note that $\overline{\mathbf{G}}$ is a 6×6 tensor, and thus it can be decomposed into electric and magnetic terms as follows:

$$\overline{\mathbf{G}} = \begin{pmatrix} \mathbf{G}_{EE} & \mathbf{G}_{EM} \\ \mathbf{G}_{ME} & \mathbf{G}_{MM} \end{pmatrix}. \tag{40}$$

For material systems with a trivial magnetic response ($\overline{\mu} = \mathbf{1}_{3\times 3}$, $\overline{\xi} = \overline{\zeta} = 0$) the electric part of the Green's function $\mathbf{G}_{EE}$, is related to the standard electric Green's function $\mathcal{G}$ (i.e., the solution of $\nabla \times \nabla \times \mathcal{G} - \frac{\omega^2}{c^2} \overline{\varepsilon} \cdot \mathcal{G} = \mathbf{1}_{3\times 3}\delta(\mathbf{r}-\mathbf{r}_0)$) as $\mathbf{G}_{EE} = i\omega\mu_0 \mathcal{G}$.

By expanding $\overline{\mathbf{G}} \cdot \hat{\mathbf{u}}_i = \sum_n \mathbf{f}_n(\mathbf{r}) c_n$ into electromagnetic modes (with $\hat{\mathbf{u}}_i$ a generic unity six-vector) and using Eq. (37) we find that [6, 27]:

$$\overline{\mathbf{G}}(\mathbf{r},\mathbf{r}_0,\omega) = \frac{i}{2} \sum_n \frac{1}{\omega_n - \omega} \mathbf{f}_n(\mathbf{r}) \otimes \mathbf{f}_n^*(\mathbf{r}_0). \tag{41}$$

It is underlined that the electromagnetic modes are normalized as in Eq. (33), and that the modal expansion of the Green's function applies only in the limit of vanishing material loss.

The summation in the right-hand side of (41) is over all modes, i.e., modes with positive, negative and zero frequencies. Using the completeness relation (34) the sum can be restricted to "transverse" modes with $\omega_n \neq 0$, so that $\overline{\mathbf{G}} = \frac{i}{2} \sum_{\omega_n \neq 0} \frac{\omega_n}{(\omega_n - \omega)\omega} \mathbf{f}_n(\mathbf{r}) \otimes \mathbf{f}_n^*(\mathbf{r}_0) - \frac{i}{\omega} \delta(\mathbf{r}-\mathbf{r}_0) \mathbf{M}_\infty^{-1}(\mathbf{r}_0)$. Furthermore, the reality of the electromagnetic fields implies that modes with positive and negative frequencies are linked by complex conjugation. Thus, the Green's function expansion can be restricted to modes with positive frequencies as follows [6, 27, 28]:

$$\overline{\mathbf{G}} = \sum_{\omega_n>0} \frac{i\omega_n}{2\omega}\left[\frac{1}{\omega_n-\omega}\mathbf{f}_n(\mathbf{r})\otimes\mathbf{f}_n^*(\mathbf{r}_0) + \frac{1}{\omega_n+\omega}\mathbf{f}_n^*(\mathbf{r})\otimes\mathbf{f}_n(\mathbf{r}_0)\right] - \frac{i}{\omega}\delta(\mathbf{r}-\mathbf{r}_0)\mathbf{M}_\infty^{-1}(\mathbf{r}_0) \quad (42)$$

We emphasize that this result applies to arbitrary bianisotropic and possibly nonreciprocal inhomogeneous material structures. Furthermore, some of the ideas described here can be readily extended to spatially dispersive materials [5, 15, 29].

The following useful property can be derived by replacing $\omega \to \omega + i0^+$ and using $\frac{1}{\omega+i0^+} = \mathrm{PV}\frac{1}{\omega} - i\pi\delta(\omega)$ with PV the principal value operator [30]:

$$\begin{aligned}&\left[(-i\omega\overline{\mathbf{G}})(\mathbf{r},\mathbf{r}_0) - (-i\omega\overline{\mathbf{G}})^\dagger(\mathbf{r}_0,\mathbf{r})\right]_{\omega+0^+ i}\\ &= i\omega\pi\sum_{\omega_n>0}\left[\delta(\omega-\omega_n)\mathbf{f}_n(\mathbf{r})\otimes\mathbf{f}_n^*(\mathbf{r}_0) + \delta(\omega+\omega_n)\mathbf{f}_n^*(\mathbf{r})\otimes\mathbf{f}_n(\mathbf{r}_0)\right]\end{aligned} \quad (43)$$

To conclude, we note that from (34) and (41) the time-domain the Green's function satisfies [27]:

$$\begin{aligned}\overline{\mathbf{G}}(\mathbf{r},\mathbf{r}_0,t) &= -u(t)\frac{1}{2}\sum_n e^{-i\omega_n t}\mathbf{f}_n(\mathbf{r})\otimes\mathbf{f}_n^*(\mathbf{r}_0)\\ &= -u(t)\frac{1}{2}\sum_{\omega_n\neq 0}\left(e^{-i\omega_n t}-1\right)\mathbf{f}_n(\mathbf{r})\otimes\mathbf{f}_n^*(\mathbf{r}_0) - \delta(\mathbf{r}-\mathbf{r}_0)\mathbf{M}_\infty^{-1}(\mathbf{r}_0)u(t)\end{aligned} \quad (44)$$

## VIII. Positive and negative frequency components of the Green's function

From the previous section, the Green's function can be decomposed as [28]:

$$\overline{\mathbf{G}} = \overline{\mathbf{G}}^+ + \overline{\mathbf{G}}^- - \frac{i}{\omega}\delta(\mathbf{r}-\mathbf{r}_0)\mathbf{M}_\infty^{-1}(\mathbf{r}_0), \quad (45)$$

with $\overline{\mathbf{G}}^+, \overline{\mathbf{G}}^-$ the positive and negative frequency components,

$$-i\omega \overline{\mathbf{G}}^{+} = \sum_{\omega_n>0} \frac{\omega_n}{2} \frac{1}{\omega_n - \omega} \mathbf{f}_n(\mathbf{r}) \otimes \mathbf{f}_n^*(\mathbf{r}_0), \quad -i\omega \overline{\mathbf{G}}^{-} = \sum_{\omega_n>0} \frac{\omega_n}{2} \frac{1}{\omega_n + \omega} \mathbf{f}_n^*(\mathbf{r}) \otimes \mathbf{f}_n(\mathbf{r}_0), \quad (46)$$

with poles in the $\text{Re}\{\omega\} > 0$ ($\text{Re}\{\omega\} < 0$) semi-spaces, respectively.

Interestingly, the components $\overline{\mathbf{G}}^{+}, \overline{\mathbf{G}}^{-}$ are relevant in some problems of quantum optics, e.g., they appear when studying the interaction of an atomic system with a quantized electromagnetic field in the framework of a Markov approximation [28]. In practice, it is convenient to evaluate $\overline{\mathbf{G}}^{+}, \overline{\mathbf{G}}^{-}$ directly in terms of the total Green's function $\overline{\mathbf{G}}$. In what follows, we establish such a link and prove that for positive frequencies ($\omega > 0$) $\overline{\mathbf{G}}^{-}$ can be written in terms of an integral of $\overline{\mathbf{G}}$ over the imaginary axis.

To begin with, we note that $\overline{\mathbf{G}}^{-}$ is analytic for $\text{Re}\{\omega\} > 0$, and thus from the Cauchy theorem [28]:

$$\left(-i\omega \overline{\mathbf{G}}^{-}\right)\bigg|_{\omega_0} = \int_{-i\infty}^{i\infty} d\omega \frac{1}{2\pi i} \frac{1}{\omega_0 - \omega} \left(-i\omega \overline{\mathbf{G}}^{-}\right)$$
$$= \frac{1}{2\pi} \int_{-\infty}^{\infty} d\xi \frac{1}{\omega_0 - i\xi} \left(-i\omega \overline{\mathbf{G}}^{-}\right)_{\omega=i\xi}. \quad (47)$$

Here, $\omega_0$ is assumed real-valued and positive. Using again the Cauchy theorem it is clear that $0 = \frac{1}{2\pi} \int_{-\infty}^{\infty} d\xi \frac{1}{\omega_0 + i\xi} \left(-i\omega \overline{\mathbf{G}}^{-}\right)_{\omega=i\xi}$. But from Eq. (46) it is evident that $\left(-i\omega \overline{\mathbf{G}}^{-}(\mathbf{r},\mathbf{r}_0,\omega)\right)_{\omega=i\xi} = \left[\left(-i\omega \overline{\mathbf{G}}^{+}(\mathbf{r},\mathbf{r}_0,\omega)\right)_{\omega=i\xi}\right]^*$. Therefore, we conclude that:

$$0 = \frac{1}{2\pi} \int_{-\infty}^{\infty} d\xi \frac{1}{\omega_0 + i\xi} \left[\left(-i\omega \overline{\mathbf{G}}^{+}(\mathbf{r},\mathbf{r}_0,\omega)\right)_{\omega=i\xi}\right]^*. \quad (48)$$

Conjugating both sides of the above equation and combining it with (47) we obtain the desired result:

$$\left(-i\omega\overline{\mathbf{G}}^-(\mathbf{r},\mathbf{r}_0,\omega)\right)\Big|_{\omega=\omega_0>0} = \frac{1}{2\pi}\int_{-\infty}^{\infty} d\xi \frac{1}{\omega_0 - i\xi}\left(-i\omega\overline{\mathbf{G}}(\mathbf{r},\mathbf{r}_0,\omega) + \delta(\mathbf{r}-\mathbf{r}_0)\mathbf{M}_\infty^{-1}(\mathbf{r}_0)\right)_{\omega=i\xi}. \quad (49)$$

As further discussed in Ref. [28], $\overline{\mathbf{G}}^-$ usually is associated with non-resonant ground state interactions, and consistent with that it is written in terms of an integral of the system Green's function over imaginary frequencies.

## IX. Application to topological photonics

Currently, there is a great interest in the use of topological methods in problems of electromagnetic wave propagation [31, 32]. Following seminal studies in the context of condensed matter theory [33-36], it has been shown that electromagnetic materials can be topologically classified by integer numbers that are insensitive to weak perturbations of the material inner structure [5, 6, 12, 13, 37, 38, 39]. Furthermore, quite remarkably it has been shown that it is possible to predict some rather fundamental wave phenomena (e.g., the propagation of scattering immune edge states) based on the topological classification of the involved materials [40-43].

The topological classification of a material is typically done relying on the eigenmodes. Both for periodic structures (e.g., photonic crystals) and electromagnetic continua (with no intrinsic periodicity) the eigenmodes, $\mathbf{f}_{n\mathbf{k}}$, are organized in bands ($n=1,2,...$) and can be labeled by a wave vector $\mathbf{k}$. The topological numbers are found from the Berry curvature $\mathcal{F}_{\mathbf{k}}$, which is written in terms of the eigenmodes. The Berry curvature is a scalar function defined over the wave vector $\mathbf{k}$-space, and it turns out that when the $\mathbf{k}$-space is a closed-manifold with no boundary the Chern number,

$$\mathcal{C} = \frac{1}{2\pi} \iint dk_x dk_y \, \mathcal{F}_{\mathbf{k}}, \tag{50}$$

is necessarily an integer, and thereby has a topological nature [5, 12]. This result is known as the Chern theorem. The integral in the above equation is over the entire wave vector space. It is possible to introduce other topological invariants for subclasses of media with specific symmetries [35, 38].

The standard definition of the Berry potential in condensed-matter systems [36] assumes that the eigenfunctions are associated with the spectrum of a Hermitian operator. Thereby, the topological classification of dispersive electromagnetic material platforms must be done based on the generalized (homogeneous) problem (12). Specifically, a straightforward generalization of the ideas of condensed-matter physics gives [5, 12]:

$$\mathcal{A}_{n\mathbf{k}} = i \langle \mathbf{Q}_{n\mathbf{k}} | \partial_{\mathbf{k}} \mathbf{Q}_{n\mathbf{k}} \rangle, \tag{51}$$

where $\mathbf{Q}_{n\mathbf{k}}$ represent the envelopes of a family of eigenmodes of the generalized system ($\hat{H}_{cl}(\mathbf{r}, -i\nabla + \mathbf{k}) \cdot \mathbf{Q}_{n\mathbf{k}} = \omega_{n\mathbf{k}} \mathbf{Q}_{n\mathbf{k}}$), $\partial_{\mathbf{k}} = \frac{\partial}{\partial k_x} \hat{\mathbf{x}} + \frac{\partial}{\partial k_y} \hat{\mathbf{y}}$, and it is implicit that the wave vector is restricted to the $k_z = 0$ plane. Furthermore, the eigenmode envelopes must be normalized as $\langle \mathbf{Q}_{n\mathbf{k}} | \mathbf{Q}_{n\mathbf{k}} \rangle = 1$. It should be noted that the Berry potential is gauge dependent because the gauge transformation $\mathbf{Q}_{n\mathbf{k}} \to \mathbf{Q}_{n\mathbf{k}} e^{i\theta_{n\mathbf{k}}}$ (with $e^{i\theta_{n\mathbf{k}}}$ a smooth function) transforms the Berry potential as $\mathcal{A}_{n\mathbf{k}} \to \mathcal{A}_{n\mathbf{k}} - \partial_{\mathbf{k}} \theta_{n\mathbf{k}}$. The Berry curvature is defined in terms of the Berry potential as

$$\mathcal{F}_{\mathbf{k}} = \hat{\mathbf{z}} \cdot \nabla_{\mathbf{k}} \times \mathcal{A}_{n\mathbf{k}}, \tag{52}$$

and is manifestly gauge independent. In general, $\mathcal{A}_{n\mathbf{k}}$ cannot be defined as a smooth vector field over the entire wave vector space. Nevertheless, it is possible to cover the $\mathbf{k}$-space with patches wherein for some gauge $\mathcal{A}_{n\mathbf{k}}$ is a smooth vector field. When $\mathcal{A}_{n\mathbf{k}}$ can be globally defined as a smooth function, it is evident from Stokes theorem that the Chern number vanishes and thereby that the material is topologically trivial. Hence, a non-zero Chern number indicates an obstruction to the application of the Stokes theorem [36]. Furthermore, it is possible to show that the Chern number always vanishes for reciprocal materials [5, 12].

Interestingly, the Berry potential can be expressed in terms of the electromagnetic component ($\mathbf{f}_{n\mathbf{k}}$) of the eigenmodes. To prove this, we use Eq. (9) to obtain:

$$\frac{1}{2}\mathbf{Q}_{n\mathbf{k}}^* \cdot \mathbf{M}_g(\mathbf{r}) \cdot i\partial_\mathbf{k} \mathbf{Q}_{n\mathbf{k}} = \frac{1}{2}\left[\mathbf{f}_{n\mathbf{k}}^* \cdot \mathbf{M}_\infty \cdot i\partial_\mathbf{k} \mathbf{f}_{n\mathbf{k}} + \mathbf{f}_{n\mathbf{k}}^* \cdot \sum_\alpha \frac{\left|\omega_{p,\alpha}\right|}{\left(\omega_{n\mathbf{k}} - \omega_{p,\alpha}\right)} \mathbf{A}_\alpha^2 \cdot i\partial_\mathbf{k}\left(\frac{1}{\omega_{n\mathbf{k}} - \omega_{p,\alpha}} \mathbf{f}_{n\mathbf{k}}\right)\right].$$

(53)

Integrating both sides of the equation over the volume $V$, noting that the Berry potential is necessarily real-valued, and using Eq. (7) it is possible to write [5, 12, 13]:

$$\mathcal{A}_{n\mathbf{k}} = \int_V d^3\mathbf{r}\, \mathrm{Re}\left\{i\mathbf{f}_{n\mathbf{k}}^* \cdot \frac{1}{2}\frac{\partial}{\partial \omega}(\omega\mathbf{M})_{\omega_{n\mathbf{k}}} \cdot \partial_\mathbf{k} \mathbf{f}_{n\mathbf{k}}\right\}. \tag{54}$$

The constraint $\langle \mathbf{Q}_{n\mathbf{k}} | \mathbf{Q}_{n\mathbf{k}} \rangle = 1$ implies that the electromagnetic fields are normalized as in Eq. (33). The application of this result to the topological classification of dispersive electromagnetic platforms is thoroughly discussed in Refs. [5-6, 42]. Furthermore, in Ref. [5-6] the result is also generalized to a subclass of spatially dispersive materials. This case is of practical importance because the application of topological concepts to electromagnetic continua typically requires the introduction of a wave vector cut-off. For more details the reader is referred to Refs. [5, 6, 42].

## X. Application to quantum optics

Modal expansions in dispersive media are of key importance in quantum optics. Indeed, the quantized electromagnetic fields in a closed cavity are expressed in terms of the electromagnetic modes. For non-dispersive structures, the time evolution of the electromagnetic field is described by a Hermitian operator and therefore the quantization procedure is straightforward: each (positive frequency) natural mode of oscillation is associated with a quantum harmonic oscillator with zero-point energy $\hbar\omega_n/2$ [4, 25, 44]. Importantly, the homogeneous (with $\mathbf{j}_g = 0$) generalized system (12) is also described by an Hermitian operator ($\mathbf{M}_g^{-1} \cdot \hat{L}$) and hence the quantization of the state vector $\mathbf{Q}$ can be done by a simple generalization of the standard procedure (see [25, Ap. A]).

To show this, first we highlight that the dynamics of the homogeneous generalized system (12) is formally equivalent to the dynamics of an infinite set of harmonic oscillators. Following the standard procedure [25], we restrict our attention to *transverse* solutions of (12), i.e., solutions $\mathbf{Q}$ that can be expanded in terms of modes with $\omega_n \neq 0$. Thus, $\mathbf{Q} = \sum_{\omega_n > 0} b_n \mathbf{Q}_n(\mathbf{r}) + b_{-n} \mathbf{Q}_{-n}(\mathbf{r})$, with the eigenmodes normalized as $\langle \mathbf{Q}_n | \mathbf{Q}_m \rangle = \delta_{n,m}$ and the inner product is defined as in Eq. (13). Here, $\mathbf{Q}_n$ is an eigenmode associated with a positive oscillation frequency $\omega_n$, and $\mathbf{Q}_{-n}$ is the corresponding eigenmode associated with the negative frequency $-\omega_n$. Furthermore, it is supposed that the electromagnetic component of $\mathbf{Q}$ is real-valued. Because of the "particle-hole" symmetry [Eq. (4)] of the Maxwell's equations, it is possible to impose that the electromagnetic component of $\mathbf{Q}_{-n}$ ($\mathbf{f}_{-n}$) is linked to that of $\mathbf{Q}_n$ ($\mathbf{f}_n$) by complex conjugation, $\mathbf{f}_{-n} = \mathbf{f}_n^*$. Then, since by hypothesis the electromagnetic component of $\mathbf{Q}$ is real-valued, it follows that $b_{-n} = b_n^*$, so that:

$$\mathbf{Q} = \sum_{\omega_n > 0} b_n \mathbf{Q}_n(\mathbf{r}) + b_n^* \mathbf{Q}_{-n}(\mathbf{r}). \tag{55}$$

The system energy is evidently equal to $H = \langle \mathbf{Q} | \mathbf{Q} \rangle$. Taking into account the normalization of the modes, we find that,

$$H = 2 \sum_{\omega_n > 0} |b_n|^2 = 2 \sum_{\omega_n > 0} \left( b_n'^2 + b_n''^2 \right), \tag{56}$$

where $b_n = b_n' + i b_n''$. In order that $\mathbf{Q}$ satisfies the equation system (12) with $\mathbf{j}_g = 0$ it is necessary that $b_n(t) = b_n(0) e^{-i\omega_n t}$, so that the real and imaginary parts of $b_n$ must satisfy $\dot{b}_n' = \omega_n b_n''$ and $\dot{b}_n'' = -\omega_n b_n'$ (the dot represents derivation in time).

Let us now introduce the variables $q_n$ and $p_n$ such that $b_n' = \frac{\omega_n}{2} \sqrt{m} \, q_n$ and $b_n'' = \frac{1}{2\sqrt{m}} p_n$, where $m$ has dimensions of mass (and can be chosen arbitrarily). Then, from the previous results it is clear that the system energy can be written as:

$$H = \sum_{\omega_n > 0} \left( \frac{1}{2} m \omega_n^2 q_n^2 + \frac{1}{2m} p_n^2 \right) \tag{57}$$

with $\dot{q}_n = \frac{1}{m} p_n = \frac{\partial H}{\partial p_n}$ and $\dot{p}_n = -m \omega_n^2 q_n = -\frac{\partial H}{\partial q_n}$. Therefore, the dynamics of the classical problem is indeed equivalent to that of an infinite set of decoupled harmonic oscillators with classical Hamiltonian given by (57).

The generalized state vector **Q** can now be quantized simply by quantizing the associated harmonic oscillators. To this end, $q_n$ and $p_n$ are promoted to operators that satisfy canonical commutation relations $[\hat{q}_n, \hat{p}_n] = i\hbar$. Furthermore, introducing the annihilation operator $\hat{a}_n = \sqrt{\frac{m\omega_n}{2\hbar}}\hat{q}_n + i\sqrt{\frac{1}{2\hbar m\omega_n}}\hat{p}_n$, it is simple to verify that the Hamiltonian of the quantized system is

$$\hat{H} = \sum_{\omega_n > 0} \hbar\omega_n \left(\hat{a}_n^\dagger \hat{a}_n + \frac{1}{2}\right), \tag{58}$$

with the quantized state vector given by (the Schrödinger picture is implicit):

$$\hat{\mathbf{Q}}(\mathbf{r}) = \sum_{\omega_n > 0} \sqrt{\frac{\hbar\omega_n}{2}} \left(\hat{a}_n \mathbf{Q}_n(\mathbf{r}) + \hat{a}_n^\dagger \mathbf{Q}_{-n}(\mathbf{r})\right). \tag{59}$$

The "hat" indicates that a given quantity must be understood as a quantum operator. As expected, $\hat{a}_n, \hat{a}_n^\dagger$ are standard creation and annihilation operators that satisfy the usual bosonic commutation relations:

$$[\hat{a}_n, \hat{a}_m] = [\hat{a}_n^\dagger, \hat{a}_m^\dagger] = 0, \qquad [\hat{a}_n, \hat{a}_m^\dagger] = \delta_{n,m}. \tag{60}$$

Since the electromagnetic fields are determined by the first element of $\hat{\mathbf{Q}}$, it is clear that Eq. (59) and $\mathbf{f}_{-n} = \mathbf{f}_n^*$ give [27, 30]:

$$\hat{\mathbf{f}} = \begin{pmatrix} \hat{\mathbf{E}} \\ \hat{\mathbf{H}} \end{pmatrix} = \sum_{\omega_n > 0} \sqrt{\frac{\hbar\omega_n}{2}} \left(\hat{a}_n \mathbf{f}_n(\mathbf{r}) + \hat{a}_n^\dagger \mathbf{f}_n^*(\mathbf{r})\right). \tag{61}$$

The condition $\langle \mathbf{Q}_n | \mathbf{Q}_m \rangle = \delta_{n,m}$ implies that the electromagnetic field natural modes are normalized as in Eq. (33). The derived formula for the quantized electromagnetic field is consistent with the results obtained with other more sophisticated approaches [45-47].

Evidently, the quantized electric displacement vector and magnetic induction are given by:

$$\hat{\mathbf{g}} = \begin{pmatrix} \hat{\mathbf{D}} \\ \hat{\mathbf{B}} \end{pmatrix} = \sum_{\omega_n>0} \sqrt{\frac{\hbar\omega_n}{2}} \left( \hat{a}_n \mathbf{g}_n(\mathbf{r}) + \hat{a}_n^\dagger \mathbf{g}_n^*(\mathbf{r}) \right), \qquad (62)$$

with $\mathbf{g}_n = \mathbf{M}(\mathbf{r},\omega_n)\cdot\mathbf{f}_n$. Using the commutation relations (60) and Eq. (32), it is simple to check that the equal time commutator of the fields satisfies:

$$\begin{aligned}
\left[\hat{\mathbf{g}}(\mathbf{r}),\hat{\mathbf{f}}(\mathbf{r}')\right] &= \sum_{\omega_n>0} \frac{\hbar\omega_n}{2}\{\mathbf{g}_n(\mathbf{r})\otimes\mathbf{f}_n^*(\mathbf{r}') - \mathbf{g}_n^*(\mathbf{r})\otimes\mathbf{f}_n(\mathbf{r}')\} \\
&= \hat{N}\cdot\frac{\hbar}{2}\sum_{\omega_n>0}\{\mathbf{f}_n(\mathbf{r})\otimes\mathbf{f}_n^*(\mathbf{r}') + \mathbf{f}_n^*(\mathbf{r})\otimes\mathbf{f}_n(\mathbf{r}')\} \\
&= \hat{N}\cdot\frac{\hbar}{2}\sum_n \mathbf{f}_n(\mathbf{r})\otimes\mathbf{f}_n^*(\mathbf{r}')
\end{aligned} \qquad (63)$$

Here, $\left[\hat{\mathbf{g}}(\mathbf{r}),\hat{\mathbf{f}}(\mathbf{r}')\right]$ represents a tensor with elements $\left[\hat{g}_i(\mathbf{r}),\hat{f}_j(\mathbf{r}')\right]$, $i,j=1,\ldots 6$. Using the completeness relation (34) we find that:

$$\left[\hat{\mathbf{g}}(\mathbf{r}),\hat{\mathbf{f}}(\mathbf{r}')\right] = \hbar\hat{N}\cdot\{\delta(\mathbf{r}-\mathbf{r}')\mathbf{M}_\infty^{-1}(\mathbf{r}')\}. \qquad (64)$$

This commutation relation applies to general bianisotropic possibly nonreciprocal and inhomogeneous material structures. Since $\mathbf{M}_\infty$ is real-valued and symmetric, the commutation relation may be rewritten as $\left[\hat{\mathbf{g}}(\mathbf{r}),\mathbf{M}_\infty(\mathbf{r}')\cdot\hat{\mathbf{f}}(\mathbf{r}')\right] = \hbar\hat{N}\cdot\{\delta(\mathbf{r}-\mathbf{r}')\mathbf{1}_{6\times 6}\}$. In particular, when the source point $\mathbf{r}'$ lies in a non-dispersive material region we find that $\left[\hat{\mathbf{g}}(\mathbf{r}),\hat{\mathbf{g}}(\mathbf{r}')\right] = \hbar\hat{N}\cdot\{\delta(\mathbf{r}-\mathbf{r}')\mathbf{1}_{6\times 6}\}$, which are the standard field commutation relations in non-dispersive bianisotropic media [25, 44]. Note that the commutation relations for the macroscopic fields are not coincident with the commutation relations in vacuum for the microscopic fields [25].

Using the developed theory, it is also straightforward to obtain the spectral density of the quantized field correlations in terms of a modal expansion. To this end, in the rest of this section we adopt the Heisenberg picture so that from (61) one finds that $\hat{\mathbf{f}}(\mathbf{r},t) = \sum_{\omega_n>0} \sqrt{\frac{\hbar\omega_n}{2}} \left( \hat{a}_n \mathbf{f}_n(\mathbf{r}) e^{-i\omega_n t} + \hat{a}_n^\dagger \mathbf{f}_n^*(\mathbf{r}) e^{+i\omega_n t} \right)$. Thus, the Fourier transform of the quantized fields satisfies:

$$\hat{\mathbf{f}}(\mathbf{r},\omega) = 2\pi \sum_{\omega_n>0} \sqrt{\frac{\hbar\omega_n}{2}} \left( \hat{a}_n \mathbf{f}_n(\mathbf{r}) \delta(\omega-\omega_n) + \hat{a}_n^\dagger \mathbf{f}_n^*(\mathbf{r}) \delta(\omega+\omega_n) \right). \tag{65}$$

For two generic scalar operators we introduce the symmetrized product as $\{\hat{A}\hat{B}\} = \frac{1}{2}(\hat{A}\hat{B} + \hat{B}\hat{A})$. Some algebra shows that the quantum vacuum expectation ($\langle \ \rangle_0$) of the tensor operator $\{\hat{\mathbf{f}}(\mathbf{r},\omega)\hat{\mathbf{f}}^\dagger(\mathbf{r}',\omega')\}$ is [30]

$$\frac{1}{(2\pi)^2} \left\langle \{\hat{\mathbf{f}}(\mathbf{r},\omega)\hat{\mathbf{f}}^\dagger(\mathbf{r}',\omega')\} \right\rangle_0$$
$$= \delta(\omega-\omega') \mathcal{E}_{0,\omega} \sum_{\omega_n>0} \frac{1}{2} \left[ \delta(\omega-\omega_n) \mathbf{f}_n(\mathbf{r}) \otimes \mathbf{f}_n^*(\mathbf{r}') + \delta(\omega+\omega_n) \mathbf{f}_n^*(\mathbf{r}) \otimes \mathbf{f}_n(\mathbf{r}') \right] \tag{66}$$

where $\mathcal{E}_{0,\omega} = \hbar|\omega|/2$ is the zero-point energy of an harmonic oscillator. With the help of Eq. (43) it is possible to write the field correlations in terms of the system Green's function [30] (note that the Green's function defined in [30] differs from the Green's function used here):

$$\frac{1}{(2\pi)^2} \left\langle \{\hat{\mathbf{f}}(\mathbf{r},\omega)\hat{\mathbf{f}}^\dagger(\mathbf{r}',\omega')\} \right\rangle_0 = \delta(\omega-\omega') \mathcal{E}_{0,\omega} \frac{-1}{2\pi} \left[ \overline{\mathbf{G}}(\mathbf{r},\mathbf{r}',\omega) + \overline{\mathbf{G}}^\dagger(\mathbf{r}',\mathbf{r},\omega) \right]_{\omega+0^+ i}. \tag{67}$$

This result corresponds to the fluctuation-dissipation theorem in the limit of a zero temperature [48]. The formula can be readily extended to the case of thermally induced fluctuations simply by replacing $\mathcal{E}_{0,\omega}$ by the energy of a harmonic oscillator at temperature $T$, i.e., $\mathcal{E}_{T,\omega} = \dfrac{\hbar\omega}{2}\coth\left(\dfrac{\hbar\omega}{2k_B T}\right)$ [48].

To conclude we note that by calculating the inverse Fourier transforms of the two operators in Eq. (66) it is found that in the time domain [30]:

$$\left\langle\left\{\hat{\mathbf{f}}(\mathbf{r},t)\hat{\mathbf{f}}^\dagger(\mathbf{r}',t)\right\}\right\rangle_T = \int_0^{+\infty} d\omega\, \mathcal{E}_{T,\omega} \sum_{\omega_n>0} \delta(\omega-\omega_n)\frac{1}{2}\left[\mathbf{f}_n(\mathbf{r})\otimes\mathbf{f}_n^*(\mathbf{r}')+\mathbf{f}_n^*(\mathbf{r})\otimes\mathbf{f}_n(\mathbf{r}')\right]. \quad (68)$$

The subscript $T$ indicates that the expectation is taken at the temperature $T$, and therefore $\mathcal{E}_{0,\omega}$ was replaced by $\mathcal{E}_{T,\omega}$.

## XI. Summary

It was proven that in the lossless case the electrodynamics of a generic inhomogeneous possibly bianisotropic and possibly nonreciprocal system may be described by an augmented state-vector whose time evolution is determined by a Hermitian operator [5, 14]. It was shown that the electromagnetic natural modes form a complete set of expansion functions and we derived different modal expansions for a generic field distribution. The modal expansions in dispersive systems are not unique because the electromagnetic degrees of freedom span a subspace of the entire Hilbert space wherein the relevant Hermitian operator is defined. Moreover, it was shown that the natural modes satisfy generalized orthogonality relations [Eqs. (33)]. The developed theory gives the system Green's function in terms of the natural modes, and has been used to find the solution of standard radiation problems [28, 29] and to characterize the light emission by moving sources [15, 27, 49].

The stored energy and canonical momentum in dispersive media were studied. We recovered a well known textbook formula for the stored energy [Eq. (20)] directly from the generalized Schrödinger-type formulation of Maxwell's equations. We also obtained an explicit formula for the canonical momentum in translation-invariant dispersive material platforms, which generalizes the findings of previous works [19, 20, 23-26].

Furthermore, it was highlighted that the Hermitian-type formulation of the dispersive Maxwell's equations enables one to extend the powerful ideas of topological photonics to a wide range of electromagnetic systems, and to introduce concepts such as the Berry potential, Berry curvature and Chern numbers which are essential to characterize the topological phases of photonic platforms [5, 12, 13]. In addition, we illustrated how the developed formalism can be applied to quantum optics. In particular, we presented a simple procedure to quantize the electromagnetic field in a generic bianisotropic and nonreciprocal closed cavity and derived the quantum correlations of the electromagnetic fields. The developed methods are useful in problems of radiative heat transport [30, 43, 50], in the study of the interactions of atomic systems with the quantized electromagnetic field [27, 28, 51], and to characterize the quantum Casimir-Polder forces [27, 28, 51].

We hope that the described methods and ideas may stimulate further studies, developments and applications of modal expansions in photonics, plasmonics and metamaterials.

**Acknowledgements:** This work is supported in part by Fundação para a Ciência e a Tecnologia grant number PTDC/EEI-TEL/4543/2014 and by Instituto de Telecomunicações under project UID/EEA/50008/2017.